# Making sense of noise: introducing students to stochastic processes in order to better understand biological behaviors

**Michael W. Klymkowsky**
Molecular, Cellular & Developmental Biology, University of Colorado Boulder, Boulder CO. 80309.  (klym@colorado.edu : ORCID: 0000-0001-5816-9771)

**Abstract**: Biological systems are characterized by the ubiquitous roles of weak, that is, non-covalent molecular interactions, small, often very small, numbers of specific molecules per cell, and Brownian (thermal) motion. These combine to produce stochastic behaviors at all levels from the molecular and cellular to the behavioral. That said, students are rarely introduced to the ubiquitous role of stochastic processes in biological systems, and how they produce unpredictable behaviors. Here I present the case that they need to be and provide some suggestions as to how it might be approached.

**Background:** Three recent events combined to spur this reflection on stochasticity in biological systems, how it is taught, and why it matters. The first was an article describing an approach to introducing students to homeostatic processes in the context of the bacterial lac operon (Booth et al., 2022), an adaptive gene regulatory system controlled in part by stochastic events. The second was in-class student responses to the question, why do interacting molecules "come back apart", that is dissociate from one another. Finally, there is the increasing attention paid to what are presented as deterministic genetic factors, as illustrated by Kathryn Harden's "The Genetic Lottery: Why DNA matters for social equality" (Harden, 2021). Previous work suggests that students, and perhaps some instructors (see Garvin-Doxas and Klymkowsky, 2008; Klymkowsky et al., 2016; 2010), find the ubiquity, functional roles, and implications of stochastic, that is processes based on unpredictable individual events, difficult to recognize and apply. Given their practical and philosophical implications, it seems essential to introduce students to stochasticity early in their educational journeys.

**What is stochasticity and why is it important for understanding biological systems?** To start, judging by word choice in Wikipedia and many scientific texts and textbooks, there seems to be widespread confusion between the differences in meaning and implication of the terms random, deterministic, and stochastic. A truly random process (essentially a miracle) would be both unpredictable and display no overall pattern at the level of individual events, large populations of events, or over time. As an example, while the decay of a single radioactive atom is not random it is unpredictable–(the basis of the Schrödinger's cat thought experiment), it is lawful. Given a large enough population of radioactive atoms, we can accurately predict the time at which 50% will have decayed - the decay of an individual atom is stochastic, not random. In biological systems, stochasticity results when intrinsically unpredictable events impact components of a system, altering the system's behavior. There are a number of drivers of stochastic behaviors. Perhaps the most obvious, and certainly the most ubiquitous in biological systems is thermal motion. The many molecules within a solution (or a cell) are moving, they have kinetic energy – the energy of motion and mass. The exact momentum of each molecule cannot, however, be accurately and completely characterized without perturbing the system (echos of Heisenberg's uncertainty principle). Given the impossibility of completely characterizing the system, we are left uncertain as to the state of the system's components, who is bound to whom, going forward.

Through collisions energy is exchanged between molecules. A number of chemical processes are driven by the energy delivered through such collisions. Think about a typical chemical reaction. In the course of the reaction, atoms are rearranged – bonds are broken (a process that requires energy) and bonds are formed (a process that releases energy). Many (most) of the chemical reactions that occur in biological systems require catalysts to bring their required activation energies into the range available, through thermal collisions, within the cell.[1]

What makes the impact of thermal motion even more critical for biological systems is that many (most) regulatory interactions and macromolecular complexes, the molecular machines discussed by Alberts (1998) are based on relatively weak, non-covalent surface-surface interactions between or within molecules. Such interactions are central to most regulatory processes, from the activation of signaling pathways to the control of gene expression. The specificity and stability of these non-covalent interactions, which include those involved in determining the three-dimensional structure of macromolecules, are directly impacted by thermal motion, and so by temperature – one reason controlling body temperature is important.

So why are these interactions stochastic and why does it matter? A signature property of a stochastic process is that while it may be predictable when large numbers of atoms, molecules, or interactions are considered, the behaviors of individual atoms, molecules, and interactions are not. A classic example, arising from factors intrinsic to the atom, is the decay of radioactive isotopes. While the half-life of a large enough population of a radioactive isotope is well defined, when any particular atom will decay is, in current theory, unknowable, a concept difficult for students (see Hull and Hopf, 2020). This is the reason we cannot accurately predict whether Schrödinger's cat is alive or dead.[2] The same behavior applies to the binding of a regulatory protein to a specific site on a DNA molecule and its subsequent dissociation: predictable in large populations, not-predictable for individual molecules. The situation is exacerbated by the fact that biological systems are composed of cells and cells are, typically, small, and so contain relatively few molecules of each type (see Milo and Phillips, 2015). There are typically one or two copies of each gene in a cell, and these may be different from one another (when heterozygous). The expression of any one gene depends upon the binding of specific proteins, transcription factors, that act to activate or repress gene expression. In contrast to a number of other cellular proteins, "as a rule of thumb, the concentrations of such transcription factors are in the nM range, corresponding to only 1-1000 copies per cell in bacteria or $10^3$-$10^6$ in mammalian cells" (Milo and Phillips, 2015). Moreover, while DNA binding proteins bind to specific DNA sequences with high affinity, they also bind to DNA "non-specifically" in a largely sequence independent manner with low affinity. Given that there are many more non-specific (non-functional) binding sites in the DNA than functional ones, the effective concentration of a particular transcription factor can be significantly lower than its total cellular concentration would suggest. For example, in the case of the lac repressor of the bacterium *Escherichia coli* (discussed further below), there are estimated to be ~10 molecules of the tetrameric lac repressor per cell, but "non-specific affinity to the DNA causes >90% of LacI copies to be bound to the DNA at locations that are not the cognate promoter site" (Milo and Phillips, 2015); at most only a few molecules are free in the cytoplasm and available to bind to specific regulatory sites. Such low affinity binding to DNA allows proteins to undergo one-dimensional diffusion, a process that can greatly speed up the time it takes for a DNA binding protein to "find" high affinity binding sites (Stanford et al., 2000; von Hippel and Berg, 1989). Most transcription factors bind in a functionally significant manner to hundreds to thousands of gene regulatory sites per cell, often with distinct binding affinities. The effective binding affinity can also be influenced by positive and negative interactions with other transcription and accessory factors, chromatin structure, and DNA modifications. Functional complexes can take time to assemble, and once assembled can initiate multiple rounds of polymerase binding and activation, leading to a stochastic phenomena known as transcriptional bursting. An analogous process occurs with RNA-dependent polypeptide synthesis (translation). The result, particularly for genes expressed at lower levels, is that stochastic (unpredictable) bursts of transcription/translation can lead to functionally significant changes in protein levels (Raj et al., 2010; Raj and van Oudenaarden, 2008).

---

[1] For this discussion we ignore entropy, a factor that figures in whether a particular reaction in favorable or unfavorable, that is whether, and the extent to which it occurs.

[2] Four common misconceptions about quantum physics [link]

There are many examples of stochastic behaviors in biological systems (Honegger and de Bivort, 2018; You and Leu, 2020). Originally noted by Novick and Weiner (1957) in their studies of the lac operon, it was clear that gene expression occurred in an all or none manner. This effect was revealed in a particularly compelling study by Elowitz et al (2002) who used lac operon promoter elements to drive expression of transgenes encoding cyan and yellow fluorescent proteins (on a single plasmid) in *E. coli* (**FIG. 1**). The observed behaviors were dramatic; genetically identical cells were found to express, stochastically, one, the other, both, or neither transgene. In a recent study the probability of bacterial cell death appears to be driven by a stochastic component (Yang et al., 2023). The stochastic expression of genes and downstream effects appear to be the source of much of the phenotypic variation found in organisms with the same genotype in the same environmental conditions (Honegger and de Bivort, 2018).

Beyond gene expression, the unpredictable effects of stochastic processes can be seen at all levels of biological organization, from the biased random walk behaviors that underlie various forms of chemotaxis (e.g. Spudich and Koshland, 1976) and search behaviors in *C. elegans* (Roberts et al., 2016) and other animals (Smouse et al., 2010), the noisiness in the opening of individual neuronal voltage-gated ion channels (Braun, 2021; Neher and Sakmann, 1976), and various processes within the immune system (Hodgkin et al., 2014), to variations in the behavior of individual organisms (e.g. the leafhopper example cited by Honegger and de Bivort, 2018). Stochastic events are involved in a range of "social" processes in bacteria (Bassler and Losick, 2006). Their impact serves as a form of "bet-hedging" in populations that generate phenotypic variation in a homogeneous environment (see Symmons and Raj, 2016). Stochastic events can regulate the efficiency of replication-associated error-prone mutation repair (Uphoff et al., 2016) leading to increased variation in a population, particularly in response to environmental stresses. Stochastic "choices" made by cells can be seen as questions asked of the environment, the system's response provides information that informs subsequent regulatory decisions (see Lyon, 2015) and the selective pressures on individuals in a population (Jablonka and Lamb, 2005). Together stochastic processes introduce a non-deterministic (i.e. unpredictable) element into higher order behaviors in a range of biological systems (Murakami et al., 2017; Roberts et al., 2016).

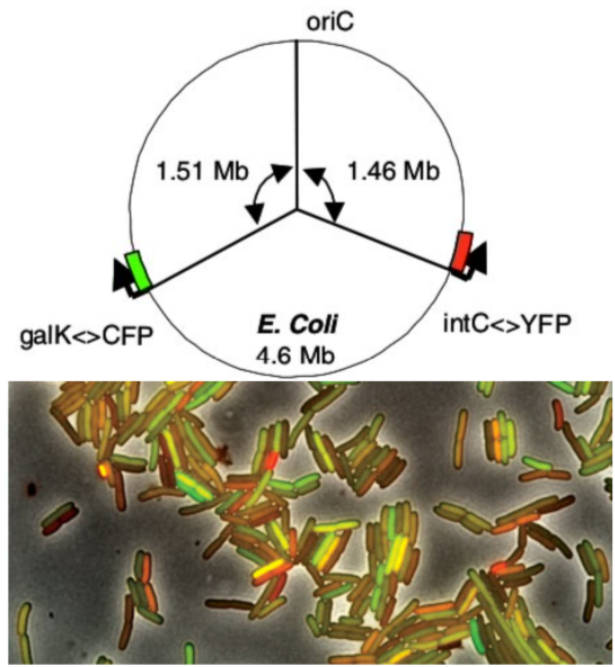

**Figure 1**: To measure sources of noise in gene expression, Elowitz et al transformed *E. coli* with a plasmid (**A**) that encoded two different fluorescent proteins whose expression was driven by lac promoters placed equidistant from the plasmid's origin of replication. (**B**) Fluorescence microscopy of transformed cells revealed the stochastic expression of the two transgenes (adapted from Elowitz et al 2002)

**Controlling stochasticity:** While stochasticity can be useful, it also needs to be controlled. Not surprisingly then there are a number of strategies for "noise-suppression", ranging from altering regulatory factor concentrations, the formation of covalent disulfide bonds between or within polypeptides, and regulating the activity of repair systems associated with DNA replication, polypeptide folding, and protein assembly via molecular chaperones and targeted degradation. For example, the identification of "cellular competition" effects has revealed that "eccentric cells" (sometimes, and perhaps unfortunately referred to as "losers") can be induced to undergo apoptosis (die) or migration in response to signals from their "normal" neighbors (Akieda et al., 2019; Di Gregorio et al., 2016; Ellis et al., 2019; Hashimoto and Sasaki, 2020; Lima et al., 2021).

**Student understanding of stochastic processes:** There is ample evidence that students (and perhaps some instructors as well) are confused by or uncertain about the role of thermal motion, that is the transfer of kinetic energy via collisions, and the resulting stochastic behaviors in biological systems. As an example, Champagne-Queloz et al (2016; 2017) found that few students, even after instruction through molecular biology courses, recognize that collisions with other molecules were responsible for the disassembly of molecular complexes. In fact, many adopt a more "deterministic" model for molecular disassembly after instruction (FIG. 2A). In earlier studies, we found evidence for a similar confusion among instructors (FIG. 2B)(Klymkowsky et al., 2010).

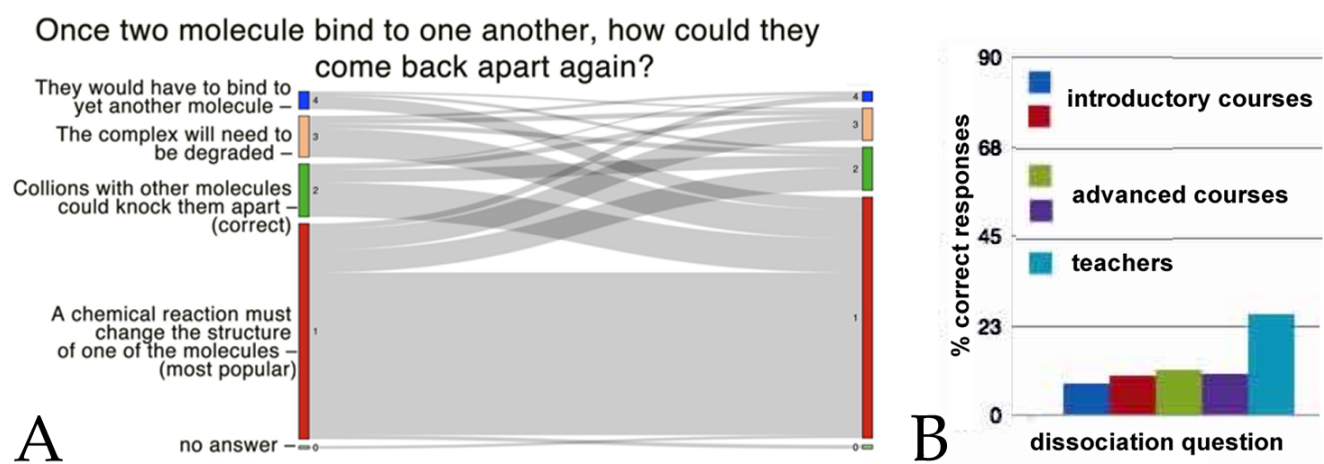

**Figure 2**: (**A**) A pre- and post-instruction analysis indicates that random molecular collisions are not recognized as causing molecular dissociation and there is little change upon instruction. The correct answer reflects the fact that molecular interactions require energy to break, and that energy comes from the transfer of energy typically through collisions with other molecules (figure from Champagne-Queloz, 2016 - used with permission). (**B**) A comparison of correct responses by different groups to this question (BCI Q18)(modified from Klymkowsky et al., 2010).

**Introducing stochasticity to students:** Given that understanding stochastic (random) processes can be difficult for many (e.g. Garvin-Doxas and Klymkowsky, 2008; Taleb, 2005), the question facing course designers and instructors is when and how best to help students develop an appreciation for the ubiquity, specific roles, and implications of stochasticity-dependent processes at all levels in biological systems. I suggest that introducing students to the dynamics of non-covalent molecular interactions, prevalent in biological systems, in the context of stochastic interactions (i.e. kinetic theory) rather than a ΔG-based approach may be useful. We can use the probability of garnering the energy needed to disrupt an interaction to present concepts of binding specificity (selectivity) and stability. Developing an understanding of the formation and disassembly of molecular interactions builds on the same logic that Albert

Einstein and Ludwig Böltzman used to demonstrate the existence of atoms and molecules and the reversibility of molecular reactions (Bernstein, 2006). Moreover, as noted by Samoilov et al (2006) "stochastic mechanisms open novel classes of regulatory, signaling, and organizational choices that can serve as efficient and effective biological solutions to problems that are more complex, less robust, or otherwise suboptimal to deal with in the context of purely deterministic systems."

> "What made the greatest impression upon the student, however, was less the technical construction of mechanics or the solution of complicated problems than the achievement of mechanics in areas which apparently had nothing to do with mechanics ... above all the kinetic theory of gases:" – A. Einstein (Bernstein)

The selectivity (specificity) and stability of molecular interactions can be understood from an energetic perspective – comparing the enthalpic and entropic differences between bound and unbound states. What is often missing from such discussions, aside from the fact of their inherent abstract nature and complexity, particularly in terms of calculating changes in entropy and exactly what is meant by energy (Cooper and Klymkowsky, 2013) is that many students enter biology classes without a robust understanding of enthalpy, entropy, or free energy (Carson and Watson, 2002). Presenting students with a molecular collision, kinetic theory-based mechanism for the dissociation of molecular interactions, may help them better understand (and apply) both the dynamics and specificity of such interactions. We can gage the strength of an interaction (the sum of the forces stabilizing the interaction) based on the amount of energy (derived from collisions with other molecules) needed to disrupt it. The implication of student responses to relevant Biology Concepts Instrument (BCI) questions (FIG 2) and course-associated activities (data not shown), as well as a number of studies in chemistry, is that few students consider the kinetic/vibrational energy delivered through collisions with other molecules (a function of temperature), as key to explaining why interactions break (see Carson and Watson, 2002 and references therein). Although this paper is 20 years old, there is little or no evidence that the situation has improved. Moreover, it appears that the conventional focus on mathematics-centered, free energy calculations in the absence of conceptual understanding may serve as an unnecessary barrier to the inclusion of more socioeconomically diverse, and under-served populations of students (Ralph et al., 2022; Stowe and Cooper, 2019).

> "*The essence of thermodynamics, however, is a study of interactions. Terms like entropy and Gibbs free energy cannot be understood as isolated entities that can be transformed into one another. If students are to use these concepts to make predictions about whether reactions can occur, they need to understand them in the context of chemical' processes'*" – Carson & Watson, 2002

**The lac operon as a context for introducing stochasticity:** Studies of the *E. coli* lac operon hold an iconic place in the history of molecular biology and are often found in introductory courses, although typically presented in a deterministic context. The mutational analysis of the lac operon helped define key elements involved in gene regulation (Jacob and Monod, 1961;

Monod et al., 1963). Booth et al (2022) used the lac operon as the context for their "modeling and simulation lesson", *Advanced Concepts in Regulation of the Lac Operon*. Given its inherently stochastic regulation (Choi et al., 2008; Elowitz et al., 2002; Novick and Weiner, 1957; Vilar et al., 2003), the lac operon is a good place to start introducing students to stochastic processes. In this light, it is worth noting that Booth et al describes the behavior of the lac operon as "leaky", which would seem to imply a low, but continuous level of expression, much as a leaky faucet continues to drip. As this is a peer-reviewed lesson, it seems likely that it reflects widely held mis-understandings of how stochastic processes are introduced to, and understood by students and instructors.

    *E. coli* cells respond to the presence of lactose in growth media in a biphasic manner, termed diauxie, due to "the inhibitory action of certain sugars, such as glucose, on adaptive enzymes (meaning an enzyme that appears only in the presence of its substrate)" (Blaiseau and Holmes, 2021). When these (preferred) sugars are depleted from the media, growth slows. If lactose is present, however, growth resumes following a delay associated with the expression of the proteins encoded by the operon that enables the cell to import and metabolize lactose. Although the term homeostatic is used repeatedly by Booth et al, the lac operon is part of an adaptive, rather than a homeostatic, system. In the absence of glucose, cyclic AMP (cAMP) levels in the cell rise. cAMP binds to and activates the catabolite activator protein (CAP), encoded for by the *crp* gene. Activation of CAP leads to the altered expression of a number of target genes, whose products are involved in adaption to the stress associated with the absence of common and preferred metabolites. cAMP-activated CAP acts as both a transcriptional repressor and activator, "and has been shown to regulate hundreds of genes in the *E. coli* genome, earning it the status of "global" or "master" regulator" (Frendorf et al., 2019). It is involved in the adaptation to environmental factors, rather than maintaining the cell in a particular state (homeostasis).

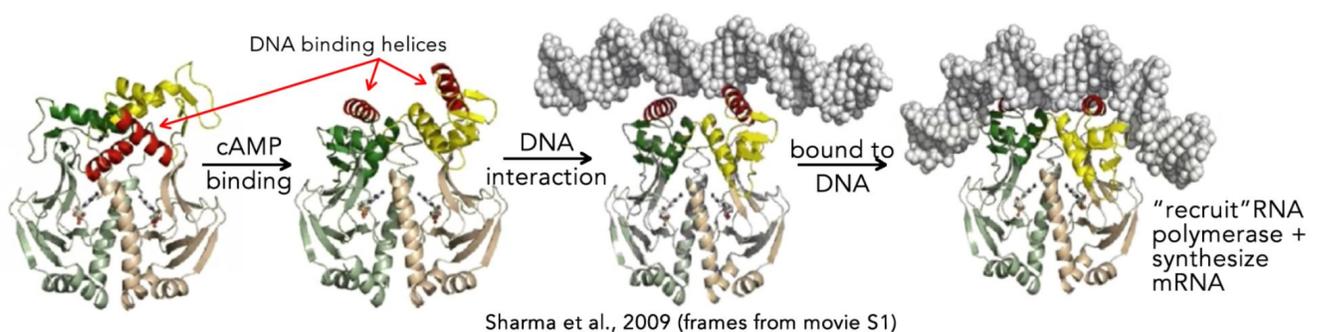

**Figure 3**: The steps in the activation of the CAP protein, upon its transition from the non-DNA binding form (left) to the allosteric transition that occurs upon the binding of cAMP, which leads to the reorientation of the molecule's DNA binding helices (adapted from Sharma et al 2009, Supplement S1)

    The lac operon is a classic polycistronic bacterial gene, encoding three distinct polypeptides: *lacZ* (β-galactosidase), *lacY* (β-galactoside permease), and *lacA* (galactoside acetyltransferase). When glucose or other preferred energy sources are present, expression of the lac operon is blocked by the inactivity of CAP. The CAP protein is a homodimer and its

binding to DNA is regulated by the binding of the allosteric effector cAMP.  cAMP is generated from ATP by the enzyme adenylate cyclase, encoded by the *cya* gene. In the absence of glucose the enyzme encoded by the *crr* gene is phosphorylated and acts to activate adenylate cyclase (Krin et al., 2002).  As cAMP levels increase, cAMP binds to the CAP protein, leading to a dramatic change in its structure (**FIG. 3**), such that the protein's DNA binding domain becomes available to interact with promoter sequences (Sharma et al., 2009).

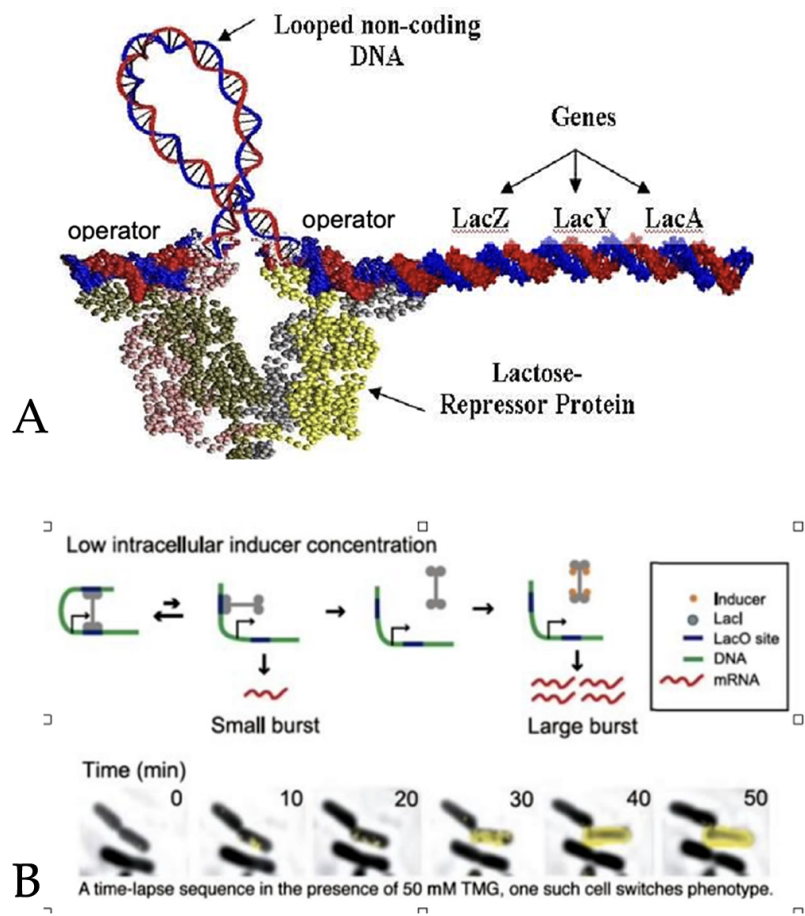

Binding of activated (cAMP-bound) CAP is not, by itself sufficient to activate expression of the lac operon because of the presence of the constitutively expressed lac repressor protein, encoded for by the *lacI* gene. The active repressor is a tetramer, present at very low levels (~10 molecules) per cell. The lac operon contains three repressor ("operator") binding sites; the tetrameric repressor can bind two operator sites simultaneously (**FIG. 4A**)(Palanthandalam-Madapusi and Goyal, 2011). In the absence of lactose, but in the presence of cAMP-activated CAP, the operon is expressed in discrete "bursts" (Novick and Weiner, 1957; Vilar et al., 2003). Choi et al (2008) found that these burst come in two types, short and long, with the size of the burst referring to the number of mRNA molecules synthesized (**FIG 4B**). The difference between burst sizes arises from the length of time that the operon's repressor binding sites are unoccupied by repressor. As noted above, the tetravalent repressor protein can bind to two operator sites at the same time. When released from one site, polymerase binding and initiation produces a small number of mRNA molecules. Persistent binding to the second site means that the repressor concentration remains locally high, favoring rapid rebinding to the operator and the cessation of transcription (RNA synthesis). When the repressor releases from both operator sites, a rarer event, it is free to diffuse away and interact (non-specifically, i.e. with low affinity) with other DNA sites in the cell, leaving the lac operator sites unoccupied for a longer period of time. The number of such non-specific binding sites greatly exceeds the number (three) of specific binding sites in the operon. The result is the synthesis of a larger "burst" (number) of mRNA molecules. The average length of time that the operator sites remain unoccupied is a function of the small number of repressor molecules present and the repressor's low but measurable non-sequence specific binding to DNA.

**Figure 4**:  Part A: Cooperativity between lac repressor protein binding to "operator" sites in the lac gene regulatory region (adapted from Palanthandalam-Madapusi & Goyal, 2011) leads (part B) to two distinct types of behavior; "small bursts" of cistron expression occur when the repressor is released from one site (but remains bound at the other), while "large bursts"  occur when the repressor disengages from both sites (adapted from Choi et al 2008).

The expression of the lac operon leads to the appearance of β-galactosidase and β-galactoside permease. An integral membrane protein, β-galactoside permease enables extracellular lactose to enter the cell while cytoplasmic β-galactosidase catalyzes its breakdown and the generation of allolactone, which binds to the lac repressor protein, inhibiting its binding to operator sites, and so removing repression of transcription. In the absence of lactose, there are few if any of the proteins (β-galactosidase and β-galactoside permease) needed to activate the expression of the lac operon, so the obvious question is how, when lactose does appear in the extracellular media, does the lac operon turn on? Booth et al and the Wikipedia entry on the lac operon (accessed 29 June 2022) describe the turn on of the lac operon as "leaky" (see above). The molecular modeling studies of Vilar et al and Choi et al (which, together with Novick and Weiner, are not cited by Booth et al) indicate that the system displays distinct threshold and maintenance concentrations of lactose needed for stable lac gene expression. The term "threshold" does not occur in the Booth et al article. More importantly, when cultures are examined at the single cell level, what is observed is not a uniform increase in lac expression in all cells, as might be expected in the context of leaky expression, but more sporadic (noisy) behaviors (**FIG. 4**). Increasing numbers of cells are "full on" in terms of lac operon expression over time when cultured in lactose concentrations above the operon's activation threshold. This illustrates the distinctly different implications of a leaky versus a stochastic process in terms of their impacts on gene expression. While a leak is a macroscopic metaphor that produces a continuous, dependable, regular flow (drips), the occurrence of "bursts" of gene expression implies a stochastic (unpredictable) process.

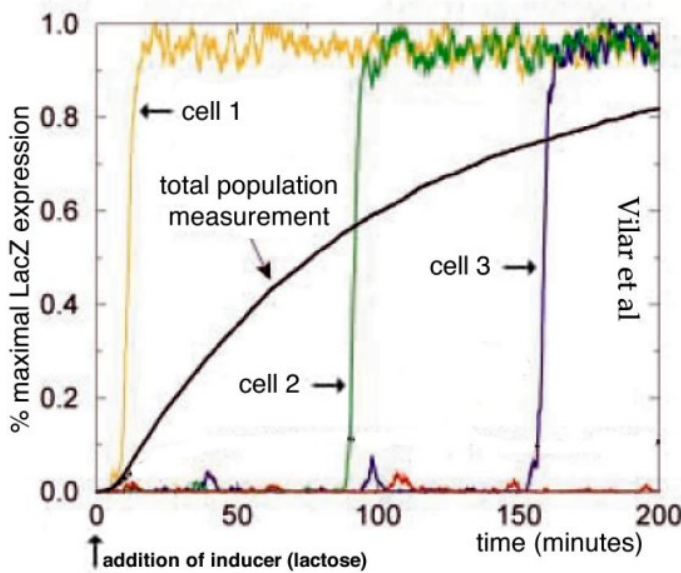

**Figure 5**: A computer simulation of the behavior, in terms of β-galactosidase (LacZ) expression of three individual *E. coli* cells (colored lines) over time in the presence of the inducer thiomethyl-β-D-galactoside. The thick black line represents the average of > 2,000 cells (Adopted from Vilar et al., 2003.

As the ubiquity and functionally significant roles of stochastic processes in biological systems becomes increasingly apparent, e.g. in the prediction of phenotypes from genotypes (Karavani et al., 2019; Mostafavi et al., 2020), helping students appreciate and understand the un-predictable, that is stochastic, aspects of biological systems becomes increasingly important. As an example, revealed dramatically through the application of single cell RNA sequencing studies, variations in gene expression between cells of the same "type" impacts organismic development and a range of behaviors. For example, in diploid eukaryotic cells is now apparent that in many cells, and for many genes, only one of the two alleles present is expressed; such "monoallelic" expression can impact a range of processes (Gendrel et al., 2014). Given that stochastic processes are often not well conveyed through conventional chemistry courses (Williams et al., 2015) or effectively integrated into, and built upon in molecular (and other) biology curricula; presenting them explicitly in introductory biology courses seems necessary and appropriate.

**Summary**: There are a number of ways that do not involve complex mathematical (or chemical energy and entropy) calculations through which introductory level biology students can be introduced into the stochastic features of biological systems, including the complex relationship between genotype and phenotype.[3] These approaches can help students appreciate (and be immunized against) the increasingly popular (apparently) illusion of genetic determinism, as illustrated by Harden's (2021) Genetic Lottery, reviewed by Feldman & Riskin (2022) and Coop & Przeworski (2022).

> "In a now classic result (Gärtner, 1990), 30 years of inbreeding experiments on laboratory mice and rats in shared environments eliminated only 20–30% of observed variance in a number of phenotypes. The remaining 70–80% was referred to as the 'intangible variance'."
> – Honegger & de Bivort, 2018

It may also help make sense of discussions of whether humans (and other organisms) have "free will". Clearly the situation is complex. From a scientific perspective we are analyzing systems without recourse to non-natural processes. At the same time, "Humans typically experience freely selecting between alternative courses of action" (Maoz et al., 2019a; see also Maoz et al., 2019b). It seems possible that recognizing the intrinsically unpredictable nature of many biological processes (including those of the central nervous system) may lead us to conclude that whether or not free will exists is in fact a non-scientific, unanswerable (and perhaps largely meaningless) question.

**Acknowledgment:** Thanks to Melanie Cooper and Nick Galati for taking a look and Chhavinder Singh for getting it started. A earlier version of this essay appear on the bioliteracy blog.

---

[3] In the context of genetics and evolutionary mechanisms, the process of genetic drift would also seem to be an appropriate context within which to introduce stochastic processes. Students consider behavior of systems as a function of population size (see https://youtu.be/B5M_C8gBvYo). Recently we have discovered a new genetic drift applet here.